\documentclass{appolb}
\usepackage{array}
\usepackage{cite}
\usepackage{color}
\usepackage{graphicx}
\begin{document}
\title{{\bf $\sigma(500)$ resonance pole positions as function of $m_\pi$: \\
analysis with a unitary coupled-channel model}\thanks
{Presented at ``Excited QCD 2024'',
Benasque, Spain, 14--20 Jan.\ 2024.}}
\author{
George Rupp\address{Centro de F\'{\i}sica Te\'{o}rica de Part\'{\i}culas,
Instituto Superior T\'{e}cnico \\ Universidade de Lisboa,
P-1049-001, Portugal}
}
\maketitle
\begin{abstract}
Resonance pole positions of the $f_0(500)$ alias $\sigma(500)$ meson are
computed and plotted as a continuous function of pion mass in the framework of
a unitary and analytic coupled-channel model for scalar mesons as dynamical
$q\bar{q}$ states. The $\sigma$ is described with a light and a strange
$q\bar{q}$ seed, mixing with each other mainly through the common $\pi\pi$,
$K\bar{K}$, and $\eta\eta$ meson-meson channels. The few model parameters are
fitted to experimental $S$-wave $\pi\pi$ phase shifts up to 1~GeV, yielding,
in the case of the physical pion mass, resonance poles at $(460-i222)$~MeV for
the $\sigma(500)$ and $(978-i37)$~MeV for the $f_0(980)$. Resonance,
bound-state, and virtual-state pole trajectories are shown as a function of
$m_\pi$ running from 139.57~MeV to 1~GeV. These are compared to recent lattice
QCD computations that use interpolating fields corresponding to the model's
channels, i.e., for a few discrete $m_\pi $values.
\end{abstract}
\section{Introduction: light scalar mesons as dynamical \boldmath$q\bar{q}$
states}
The light scalar mesons $f_0(500)$ (alias $\sigma(500)$), $f_0(980)$,
$K_0^\star(700)$ (alias $\kappa(700)$), and $a_0(980)$ \cite{PDG22} have been
haunting theorists and experimentalists for more than the past half-century
(see e.g.\ the minireview in Ref.~\cite{RB18}). Their PDG experimental status
only stabilised in 2018, with the above name and mass assignments, thus
confirming that they form a complete $SU(3)$-flavour nonet, as already
proposed by R.~L.~Jaffe in 1977 \cite{Jaffe77}. However, in his approach the
light scalars were described as ground-state $q^2\bar{q}^2$ (``tetraquark'')
states, owing their low masses to a very large attractive colour-hyperfine
interaction in the framework of the MIT Bag Model \cite{MIT74}, resulting in
the predictions $\epsilon(650)$, $S^\star(1100)$, $\kappa(900)$, and
$\delta(1100)$, with $\epsilon$, $S^\star$, and $\delta$ being the then used
PDG names of the $\sigma(500)$, $f_0(980)$, and $a_0(980)$, respectively. But
Jaffe himself already warned \cite{Jaffe77,RB17} that the usual accuracy in
calculating $q\bar{q}$ masses should not be expected in his $q^2\bar{q}^2$
model for the light scalars, which ignores decay processes for two very broad
states ($\epsilon$/$\sigma$ and $\kappa$). Moreover, he emphasised
\cite{Jaffe77,RB17} that his tetraquarks can simply fall apart into two
mesons, not requiring the creation of an additional $q\bar{q}$ pair. In other
words, such decay strengths are of order $N_c^0$, instead of $N_c^{-1}$ for
ordinary mesons. For more discussion on other non-exotic tetraquark candidates
and tetraquarks (besides pentaquarks) in lattice simulations, see the reviews
in Refs.~\cite{BR20} and \cite{Bicudo23}, respectively.

\begin{figure}[!t]
\centering
\includegraphics[trim = 43mm 150mm 20mm 46mm,clip,width=12.6cm,angle=0]
{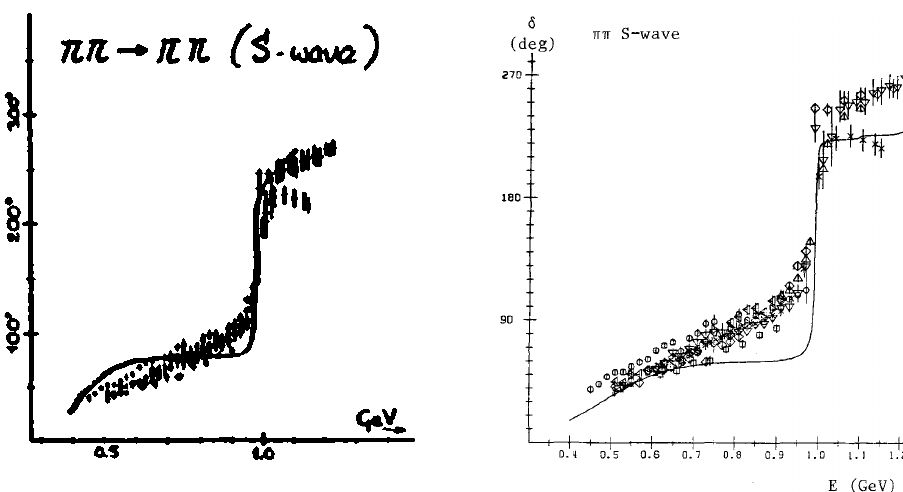}
\caption{$S$-wave $\pi\pi$ phase shifts (full lines) from Refs.~\cite{Eef83}
(left) and \cite{Beveren86} (right).}
\label{pipiS}
\end{figure}

In an alternative approach, my co-authors and I showed \cite{Eef83} that the
$\sigma(500)$ can be dynamically generated as a $q\bar{q}$ resonance in a
coupled-channel quark-meson model, as an extra state emerging from the
$\pi\pi$ continuum besides a regular $P$-wave $q\bar{q}$ scalar around
1.3~GeV. Moreover, even a rough description of the $S$-wave $\pi\pi$ phase
shifts resulted \cite{Eef83} from a model \cite{Beveren83} calculation with
unchanged parameters (see Fig.~\ref{pipiS}, left-hand plot). Application,
without any new fit, of the same model to the complete scalar nonet
\cite{Beveren86} yielded resonance pole positions that are still within
current PDG bounds \cite{PDG22}, viz.\ $f_0(470-i208)$, $f_0(994-i20)$,
$K_0^\star(727-i263)$, and $a_0(968-i28)$, employing here their modern PDG
designations. (Also see Fig.~\ref{pipiS}, right-hand plot, for a crude
reproduction \cite{Beveren86} of the $S$-wave $\pi\pi$ phases.)
A much more recent yet largely equivalent unitary multichannel model
\cite{BR21}, now formulated in momentum space in the framework of 
the so-called Resonance-Spectrum Expansion (RSE) \cite{BR09}, was fitted to 
$S$-wave $\pi\pi$ phase shifts up to 1.6~GeV and the $a_0(980)$ line shape,
resulting in $f_0(500)$, $f_0(980)$, $f_0(1370)$, $a_0(980)$, 
and $a_0(1450)$ pole positions compatible with their PDG
\cite{PDG22} entries.

\begin{figure}[!t]
\centering
\includegraphics[trim = 43mm 212mm 7mm 45mm,clip,width=12.6cm,angle=0]
{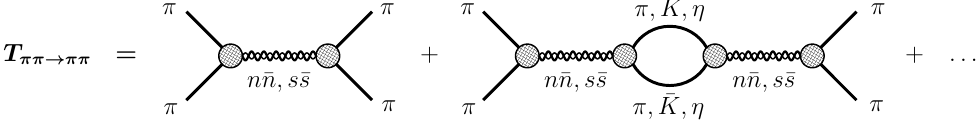}
\caption{Graphical representation of the model's $T_{\pi\pi\to\pi\pi}$
amplitude.}
\label{T_pipi}
\end{figure}

The present study deals with the $\sigma(500)$ showing up as a bound state
(BS), virtual (bound) state (VBS), or resonance, depending on the model
choice of an unphysical large pion mass. The goal is to compare these
predictions to the recent lattice QCD (LQCD) calculations in
Refs.~\cite{Briceno17,Briceno18,Rodas23a,Rodas23b}, in which it has not yet
been possible to obtain stable signals for the physical $m_\pi$ and only a few
larger pion masses have been used. So I focus here exclusively on real and
complex $\sigma(500)$ pole positions as a continuous function of $m_\pi$, in a
restricted version of the RSE model of Ref.~\cite{BR21}. It is limited to the 
$n\bar{n}\equiv (u\bar{u}\!+d\bar{d})/\sqrt{2}$ and $s\bar{s}$ quark-antiquark
channels coupled to the $\pi\pi$, $K\bar{K}$, and $\eta\eta$ two-meson
channels. These channels correspond to the interpolating fields used in
the LQCD computations of Refs.~\cite{Briceno18,Rodas23a,Rodas23b}, while
Ref.~\cite{Briceno17} only did not include an $\eta\eta$ interpolator. The
three RSE model parameters are fitted to the $S$-wave $\pi\pi$ phase shifts up
to 1~GeV, with the corresponding $T_{\pi\pi\!\to\!\pi\pi}$ amplitude
symbolically depicted in Fig.~\ref{T_pipi}. For further model details, see
Ref.~\cite{Rupp24}. The resulting fit to the $\pi\pi$ phases is quite good,
especially at low energies (see Ref.~\cite{Rupp24} for a plot), besides a
very reasonable $S$-wave $\pi\pi$ scattering length $a_0^0=0.211m_\pi^{-1}$.
Also, the extracted pole positions $\sigma(460\!-\!i222)$ and
$f_0(978\!-\!i37)$ are again inside PDG \cite{PDG22} limits.
In Sec.~\ref{traj} I shall present the resulting $\sigma$ pole trajectories
in the complex $E$ and $k$ planes.

\section{\boldmath$\sigma(500)$ pole trajectories as a function of
\boldmath$m_\pi$}
\label{traj}
\begin{figure}[!t]
\hspace*{-6mm}
\begin{tabular}{lr}
\includegraphics[trim = 43mm 95mm 20mm 30mm,clip,width=6.3cm,angle=0]
{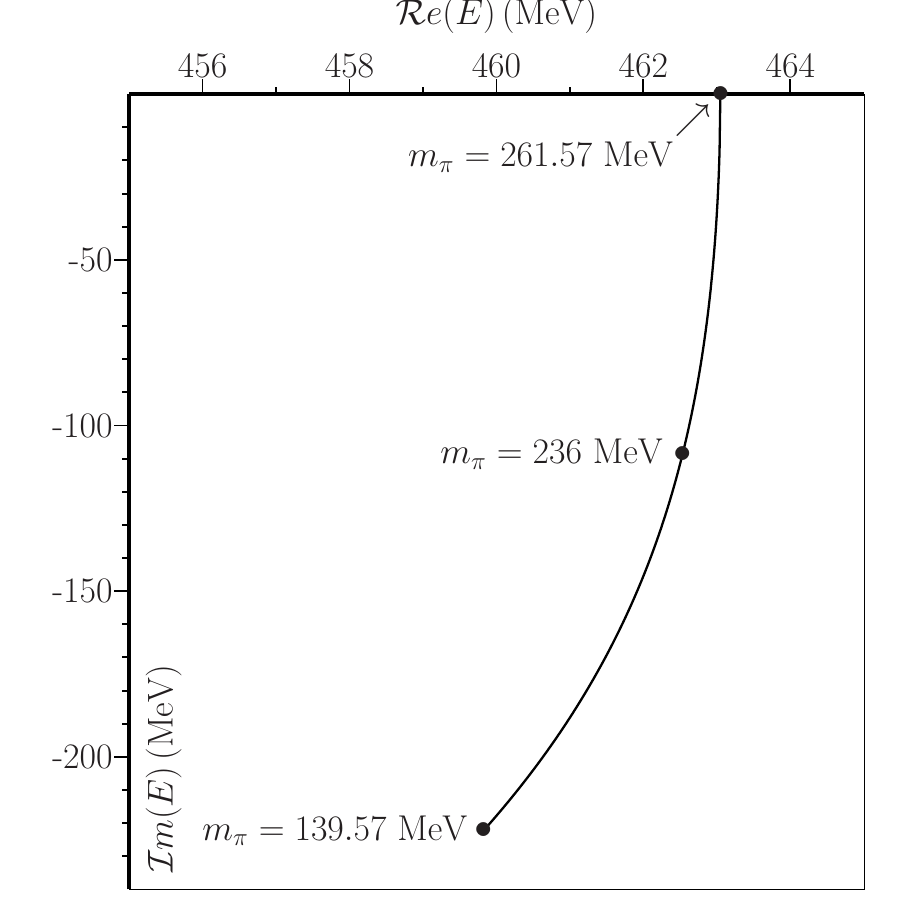}
&
\includegraphics[trim = 43mm 95mm 20mm 30mm,clip,width=6.3cm,angle=0]
{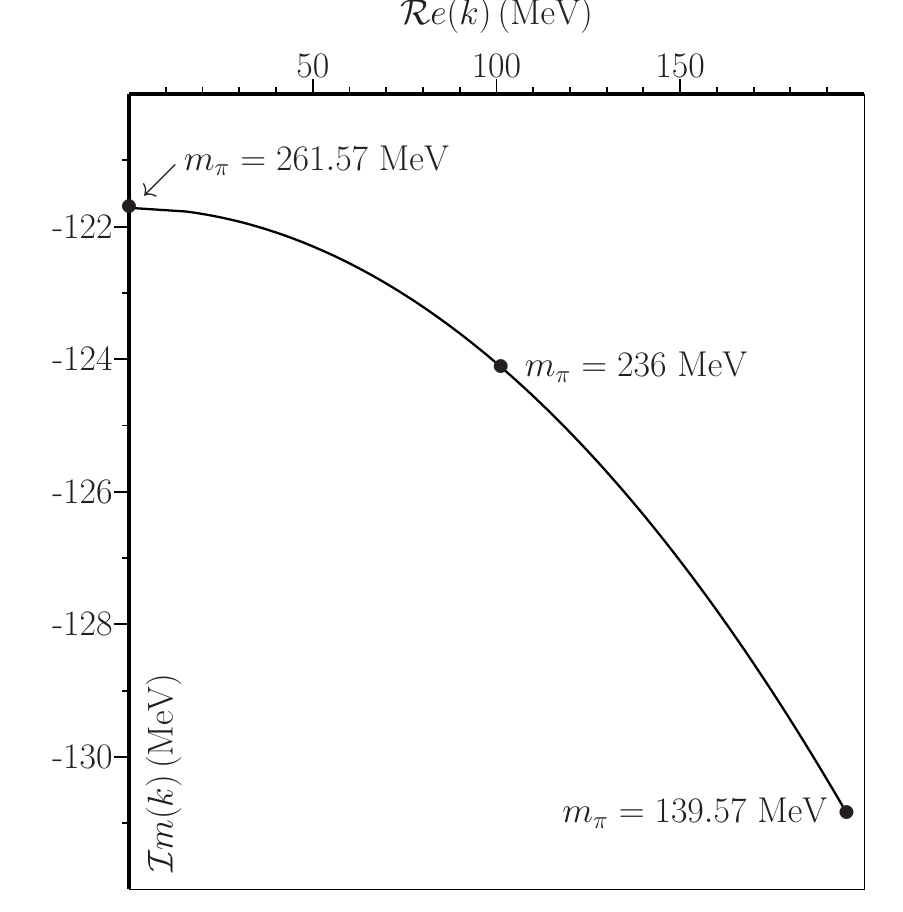}
\end{tabular}
\caption{$\sigma(500)$ resonance pole trajectories as a function of $m_\pi$.
Left: complex $E$ plane; right: complex $k$ plane. Points $m_\pi=236$~MeV:
value chosen in Ref.~\cite{Briceno17}.}
\label{Res_Ek}
\end{figure}

\begin{figure}[!b]
\hspace*{-6mm}
\begin{tabular}{lr}
\includegraphics[trim = 43mm 95mm 20mm 30mm,clip,width=6.3cm,angle=0]
{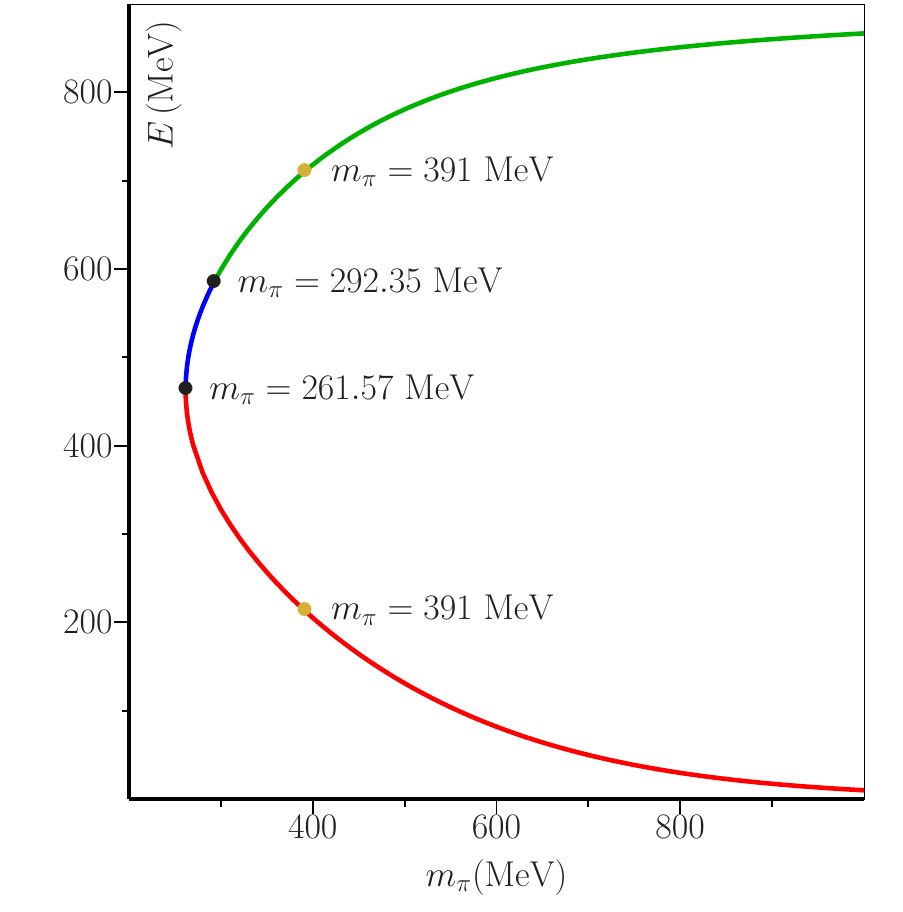}
&
\includegraphics[trim = 43mm 95mm 20mm 30mm,clip,width=6.3cm,angle=0]
{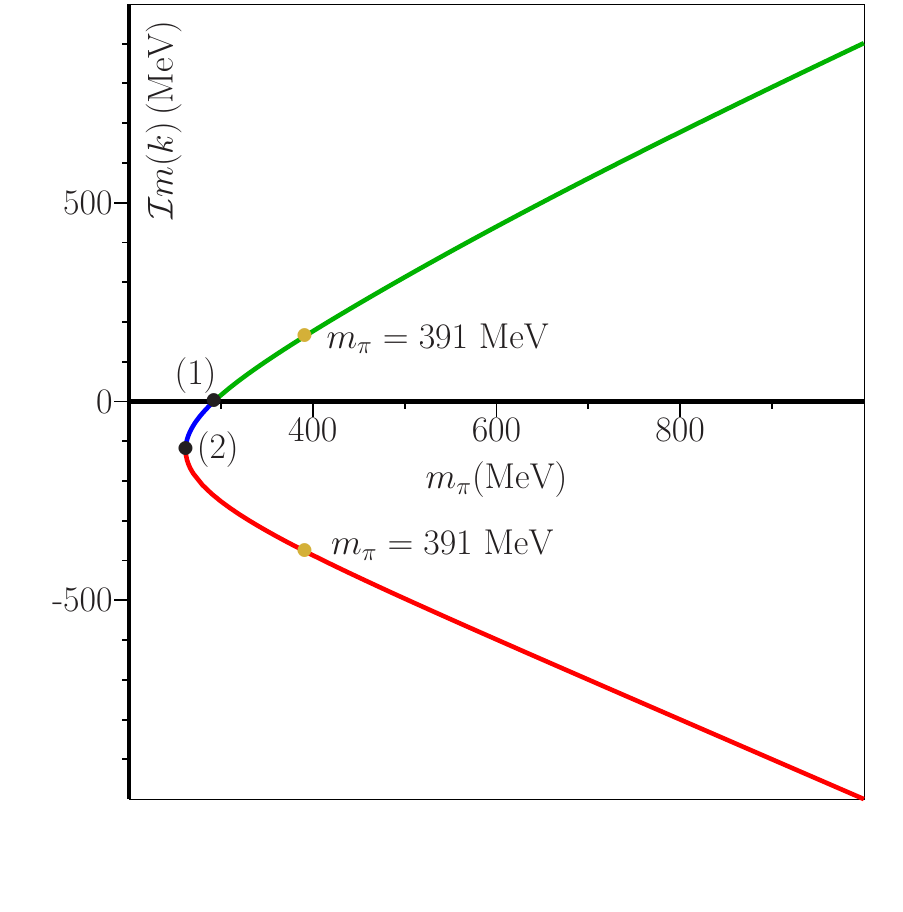}
\end{tabular}
\caption{$\sigma(500)$ real pole trajectories as a function of
$m_\pi$. Green: BS; blue: first VBS; red: second VBS.
Left: complex $E$ plane; right: complex $k$ plane. Points $m_\pi=391$~MeV:
value chosen in Refs.~\cite{Briceno17,Briceno18,Rodas23b}.}
\label{BV_Ek}
\end{figure}

Starting at the physical $m_\pi$, we see in the two plots of Fig.~\ref{Res_Ek}
how the $\sigma$ pole evolves from a regular resonance to a subthreshold one
for $m_\pi\approx231$~MeV and then ends up on the real $E$ and negative
imaginary $k$ axis. We will see below that this amounts to one complex pole
turning into a pair of real $E$ or imaginary $k$ poles, respectively. Such
subthreshold resonances are exclusive to $S$-waves (see e.g.\
Ref.~\cite{Hanhart14}). Finally, Fig.~\ref{BV_Ek} displays the real
trajectories corresponding to BS and VBS states, for
$m_\pi\geq261.57$~MeV. So these trajectories are double-valued functions of
$m_\pi$, with a left-hand VBS for each BS or right-hand VBS, in full agreement
with the $\sigma$ pole trajectories as a function of overall coupling constant
in Refs.~\cite{BR21,BR23}. For an earlier study of the $\sigma$ as a function
of $m_\pi$, in the context of unitarised chiral perturbation theory, see
Ref.~\cite{Albaladejo12}.

\section{Comparison to recent lattice QCD computations}
In Refs.~\cite{Briceno17,Briceno18,Rodas23a,Rodas23b}, LQCD computations of
the $\sigma(500)$ were presented for pion masses of 236/391, 391, 283/330,
and 239/283/330/391~MeV, respectively. In Refs.~\cite{Briceno17,Briceno18},
a BS was found for $m_\pi=758\,(4)$ and $745\,(5)$~MeV, respectively, to be
compared to 710~MeV in the unchanged present model. However, if the Kaon and
$\eta$ masses are taken at the values fixed in Ref.~\cite{Briceno18} and
a phenomenological subthreshold suppression of the $K\bar{K}$ and $\eta\eta$
is included as in Ref.~\cite{BR21}, the BS mass increases to 752~MeV. Turning
to the $\sigma$ resonance computations in Ref.~\cite{Briceno17}, widely spread
out pole positions were presented, resulting from different parametrisations
used to extrapolate the found real amplitudes into the complex plane. The
central values of these poles came out in a range of about 590--760~MeV for
the real parts and approximately 280--460~MeV for the widths, with error bars
of the order of $\pm80$~MeV. So all these pole lie clearly above the $\pi\pi$
threshold at 472~MeV, to be contrasted with the subthreshold resonance poles in
Fig.~\ref{Res_Ek} above, namely for $m_\pi$ between 231 and 261~MeV.

Still in the bound-state case, two further pion masses were explored in
Ref.~\cite{Rodas23a}, viz.\ $m_\pi=283$ and 330~MeV. From these calculations
the authors concluded that a transition from a BS to either a VBS or a
resonance below threshold occurs somewhere between these two pion masses, with
a favoured scenario of a VBS in a narrow interval of $m_\pi$ before the pole 
turns into a subthreshold resonance. This is in agreement with the
pole behavior in Fig.~\ref{BV_Ek}, with the BS $\to$ VBS transition taking
place at $m_\pi\approx292$~MeV.

Finally, in Ref.~\cite{Rodas23b} a significant improvement was obtained
regarding the $\sigma$ resonance poles for $m_\pi=239$~MeV, being equivalent
to the value $m_\pi=236$~MeV reported in Ref.~\cite{Briceno17}. This was
achieved employing dispersive methods, resulting in a range of 498--586~MeV
for the real parts of the poles and 394--506~MeV for the widths, with a 
maximum error of $\pm82$~MeV. So these results are compatible with a possible
subthreshold resonance in the case $m_\pi=239$~MeV, at least for two
parametrisations \cite{Rodas23b}. However, there is still a serious
disagreement with the model prediction of the $\sigma$ width for this pion
mass, which is about 220~MeV (see Fig.~\ref{Res_Ek}).

In conclusion, it would be very clarifying if a few further pion masses
between 239 and roughly 300~MeV could be explored in the referred LQCD
computations, in order to verify an inevitably very fast decrease of the
$\sigma$ width towards zero in this relatively narrow $m_\pi$ range.
Furthermore, I would also consider it interesting to see if some kind of
$\sigma$ resonance pole can survive if no $q\bar{q}$ interpolators are
included besides the two-meson ones $\pi\pi$, $K\bar{K}$, and $\eta\eta$.
A similar study was carried out in the LQCD calculation \cite{Padmanath15}
of the axial-vector charmonium-like meson $\chi_{c1}(3872)$ (alias $X(3872)$),
concluding that this state does not survive if no $c\bar{c}$ interpolator
is included.

\section*{Acknowledgement}
I am indebted to R.~J.~Perry, University of Barcelona, for having drawn my
attention to the very recent LQCD calculations in
Refs.~\cite{Rodas23a,Rodas23b}.

\end{document}